\documentclass[final,5p,times]{my_elsarticle}

\usepackage{epsfig,lineno}
\biboptions{sort&compress}

\begin{document}
\begin{frontmatter}

\title{Aboveground test of an advanced Li$_2$MoO$_4$ scintillating bolometer to search for neutrinoless double beta decay of $^{100}$Mo}

\author[SIGM]{T.B.~Bekker}
\author[IAS]{N.~Coron}
\author[KINR]{F.A.~Danevich}
\author[KNU]{V.Ya.~Degoda}
\author[CSNSM,Como,INFN-Bicocca]{A.~Giuliani\corref{cor1}}
\cortext[cor1]{Corresponding author. E-mail:
Andrea.Giuliani@csnsm.in2p3.fr}
\author[NIIC]{V.D.~Grigorieva}
\author[NIIC]{N.V.~Ivannikova}
\author[CSNSM,Como]{M.~Mancuso}
\author[CSNSM]{P.~de~Marcillac}
\author[KNU]{I.M.~Moroz}
\author[CEA]{C.~Nones}
\author[CSNSM]{E.~Olivieri}
\author[INFN-Bicocca]{G.~Pessina}
\author[KINR,CSNSM]{D.V.~Poda}
\author[NIIC]{V.N.~Shlegel}
\author[KINR]{V.I.~Tretyak}
\author[ICMCB]{M.~Velazquez}

\address[SIGM]{V.S.~Sobolev Institute of Geology and Mineralogy of the Siberian Branch of the RAS, 630090 Novosibirsk, Russia}
\address[IAS]{IAS, Batiment 121, UMR 8617 Universite Paris-Sud 11/CNRS, 91405 Orsay, France}
\address[KINR]{Institute for Nuclear Research, MSP 03680 Kyiv, Ukraine}
\address[KNU]{Kyiv National Taras Shevchenko University, MSP 03680 Kyiv,
Ukraine}
\address[CSNSM]{CNRS/CSNSM, Centre de Sciences Nucl$\acute{e}$aires et de Sciences de la Mati$\grave{e}$re, 91405 Orsay,
France}
\address[Como]{Dipartimento di Scienza e Alta Tecnologia dell'Universit\`a dell'Insubria, 22100 Como, Italy}
\address[INFN-Bicocca]{INFN, Sezione di Milano-Bicocca, 20126 Milano, Italy}
\address[NIIC]{Nikolaev Institute of Inorganic Chemistry, 630090 Novosibirsk, Russia}
\address[CEA]{Service de Physique des Particules, CEA-Saclay, 91191 Gif sur Yvette, France}
\address[ICMCB] {CNRS, Universite de Bordeaux, ICMCB, 33608 Pessac cedex,
France}

\begin{abstract}

Large lithium molybdate (Li$_2$MoO$_4$) crystal boules were produced by using the low thermal gradient Czochralski growth technique from deeply
purified molybdenum. A small sample from one of the boules was preliminary characterized in terms of X-ray-induced and thermally-excited luminescence. A large cylindrical crystalline element (with a size of $\oslash 40\times40$ mm) was used to fabricate a scintillating bolometer, which was operated aboveground at $\sim 15$~mK by using a pulse-tube cryostat housing a high-power dilution refrigerator. The excellent detector performance in terms of energy resolution and $\alpha$ background suppression along with preliminary positive indications on the radiopurity of this material show the potentiality of Li$_2$MoO$_4$ scintillating bolometers for low-counting experiment to search for neutrinoless double beta decay of $^{100}$Mo.

\end{abstract}


\end{frontmatter}

\section{Introduction}
\label{sec:intro}

Neutrinoless double beta ($0\nu2\beta$) decay  (see the recent reviews \cite{Cre14,Sch13,Ell12,Ver12,Giu12,Gom12,Rod11} and references therein) is a rare nuclear transition which plays a unique role in testing lepton number conservation and in understanding fundamental neutrino properties. It consists in the transformation of an even-even nucleus into a lighter isobar containing two more protons and accompanied by the emission of two electrons and no other particles. Since the total lepton number changes by two units, $0\nu2\beta$ decay is forbidden in the Standard Model, but foreseen in many of its extensions, which explain naturally the smallness of the neutrino masses. An observation of this rare process would establish that neutrinos are Majorana particles, i.e. the only fermions that coincide with their own antiparticles. The $0\nu2\beta$ decay may be induced by a plethora of mechanisms. Among them, the so-called mass mechanism -- consisting in the exchange of a virtual
 light Majorana neutrino --  occupies a special place, since its possible occurrence is directly related to the non-zero values of neutrino masses suggested by neutrino flavor oscillations.

The $0\nu2\beta$ decay distinctive signal is a monochromatic peak in the sum energy spectrum of the two emitted electrons at the $Q$-value of the transition. Given the long expected lifetime of the $0\nu2\beta$ process ($> 10^{25}$~y), searching for it requires large sources, containing tens or hundreds of kg of the isotope of interest.  Desirable detector features are high energy resolution and efficiency, and as low as possible (ideally zero) background. The choice of the nucleus to be studied is an essential element too. Among tens of possible candidates, the isotope $^{100}$Mo is one of the most promising because of its high transition energy  ($Q = 3034.40(17)$~keV~\cite{Qval}), reasonable natural isotopic abundance ($\delta  = 9.824(50)$~\%~\cite{AI}), and favorable theoretical predictions (see Ref.~\cite{Ver12,Fog2008,Bar2012} and references therein). Last but not least, molybdenum enrichment in the isotope $^{100}$Mo can be done at a reasonable price and throughput~\cite{Giu12}. The most stringent half-life limit on the $0\nu2\beta$ decay of  $^{100}$Mo was set by the NEMO3 collaboration at $T_{1/2} = 1.1\times10^{24}$~ at 90\% confidence level~\cite{Arn2014}.

\nopagebreak
\begin{table}[t]
 \caption{Properties of Li$_2$MoO$_4$ crystal scintillators}
\begin{center}
\begin{tabular}{lll}

 \hline
 Property                               & Value                 & Reference \\
  ~                                     & ~                     & ~         \\
  \hline

 Density (g/cm$^3$)                     & $3.02-3.07$           & \cite{Barinova:2010} \\

 \hline
 Melting point (K)                      & $974\pm2$             & \cite{Denielou:1975}     \\

 \hline

 Hygroscopicity                         & Weak                  & \cite{Barinova:2010} \\

 \hline

 Index of refraction                    & 1.44                  & \cite{Tkhashokov:2009} \\

 \hline
 Wavelength                  & $540$ at 85 K                     & \cite{Barinova:2010}  \\
 of maximum                           &    $590$ at 8 K      & Present work\\

emission (nm)                                     & $600$ at 85 K         & Present work \\

 \hline
 Radioactive      & ~                     & ~ \\
 contamination (mBq/kg)     & ~                     & ~ \\

 $^{40}$K                                  & $170(80)$             & \cite{Barinova:2010} \\

 $^{232}$Th                                & $\leq 0.11$           & \cite{Cardani:2013} \\

 $^{238}$U                                 & $\leq 0.09$           & \cite{Cardani:2013} \\

 \hline
 \end{tabular}
  \label{properties}
 \end{center}
 \end{table}

Scintillating bolometers look one of the most promising technique to search for $0\nu2\beta$ decay in several isotopes~\cite{Pir2006, Giu2012} thanks to high
detection efficiency (achieved in the so-called source=detector approach), excellent energy resolution (typical for the bolometric technique) and potentially very low background counting rate (obtained by exploiting their excellent $\alpha$/$\beta$ rejection power). 

There are several inorganic crystal scintillators containing molybdenum (molybdates) that can be used in principle as a basic element for a scintillating bolometer embedding $^{100}$Mo. The most promising of them are ZnMoO$_4$, CaMoO$_4$, CdMoO$_4$, PbMoO$_4$ and Li$_2$MoO$_4$. However, CdMoO$_4$ contains the $\beta$~active $^{113}$Cd, which has also a very high cross section for thermal neutron capture. A potential disadvantage of PbMoO$_4$ is that  $^{100}$Mo would be only 27\% of the total mass. For this reason, the research has focused so far on the first two compounds. Scintillating bolometers of CaMoO$_4$ (enriched in $^{100}$Mo and depleted in the radioactive $^{48}$Ca) have been successfully developed within the AMoRE collaboration~\cite{Bha12,Kim14}, while an intense activity on ZnMoO$_4$ is ongoing in the LUMINEU and LUCIFER experiments with remarkable results~\cite{Gir10,Bee12,Art14,Bee12a,Bee12b,Bee12c,Che13,Ber14}, including enriched Zn$^{100}$MoO$_4$ detectors~\cite{enr-above}.

This paper concerns the extremely promising results achieved with a lithium molybdate (Li$_2$MoO$_4$) scintillating bolometer of remarkable volume ($\sim 50$ cm$^3$). Important advantages of Li$_2$MoO$_4$ in comparison to other molybdate crystal scintillators are the highest concentration of molybdenum (55.2\% in weight), absence of natural long-living radioactive isotopes (as $\beta$~active $^{210}$Pb in PbMoO$_4$, $2\beta$~active $^{48}$Ca in CaMoO$_4$, $\beta$~active $^{113}$Cd and $2\beta$~active $^{116}$Cd in CdMoO$_4$), and comparatively easy crystal growth process. However, initial tests showed a low light yield. In this paper, we confirmed that the scintillation efficiency is the lowest among all the aforementioned molybdates, but that a well-performing light detector can provide anyway an excellent separation between $\beta$-like and $\alpha$ events, at the level required for a high sensitive $0\nu2\beta$ decay search.

\begin{figure*}
\begin{center}
\resizebox{0.38\textwidth}{!}{\includegraphics{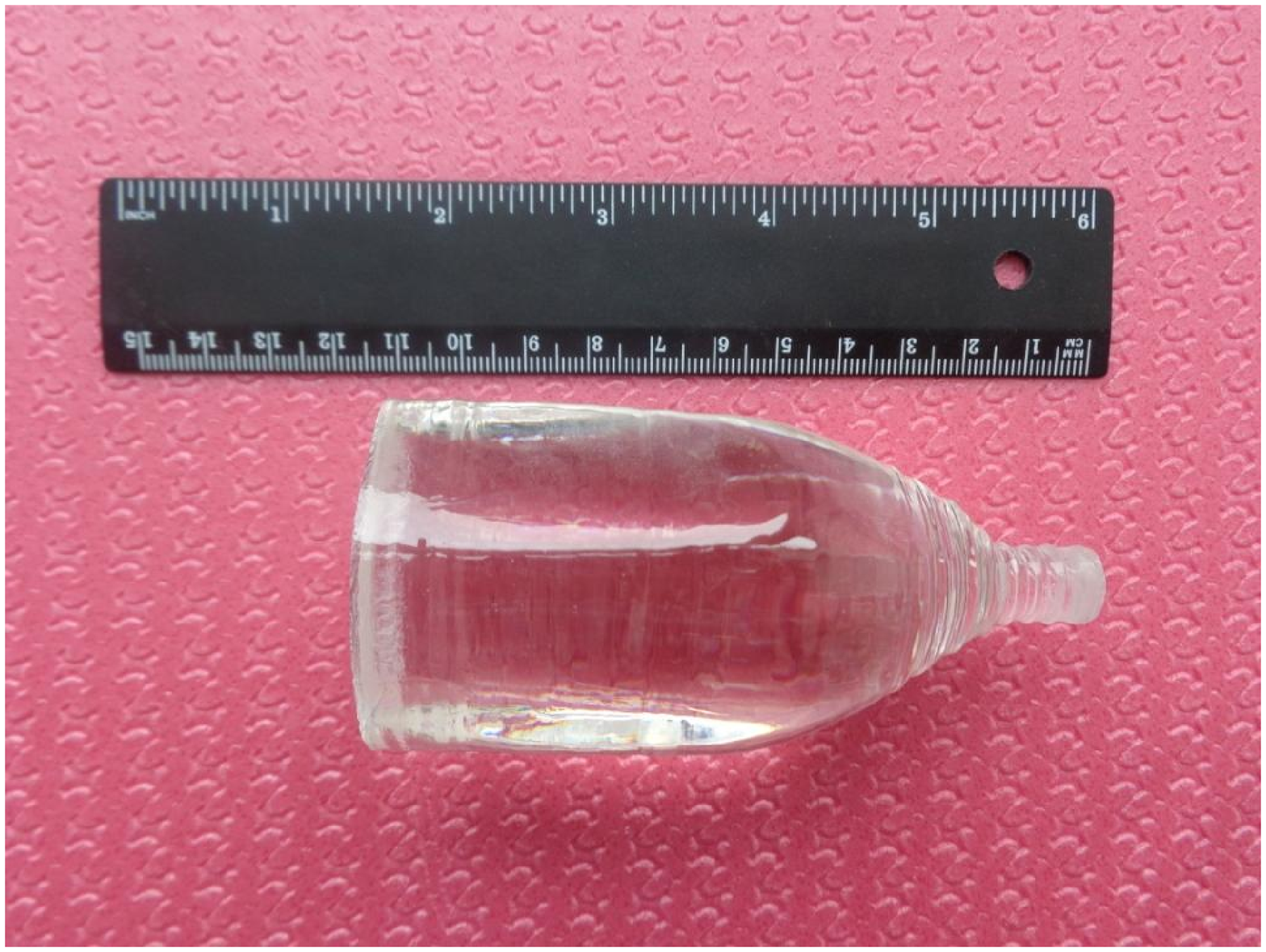}}
\resizebox{0.38\textwidth}{!}{\includegraphics{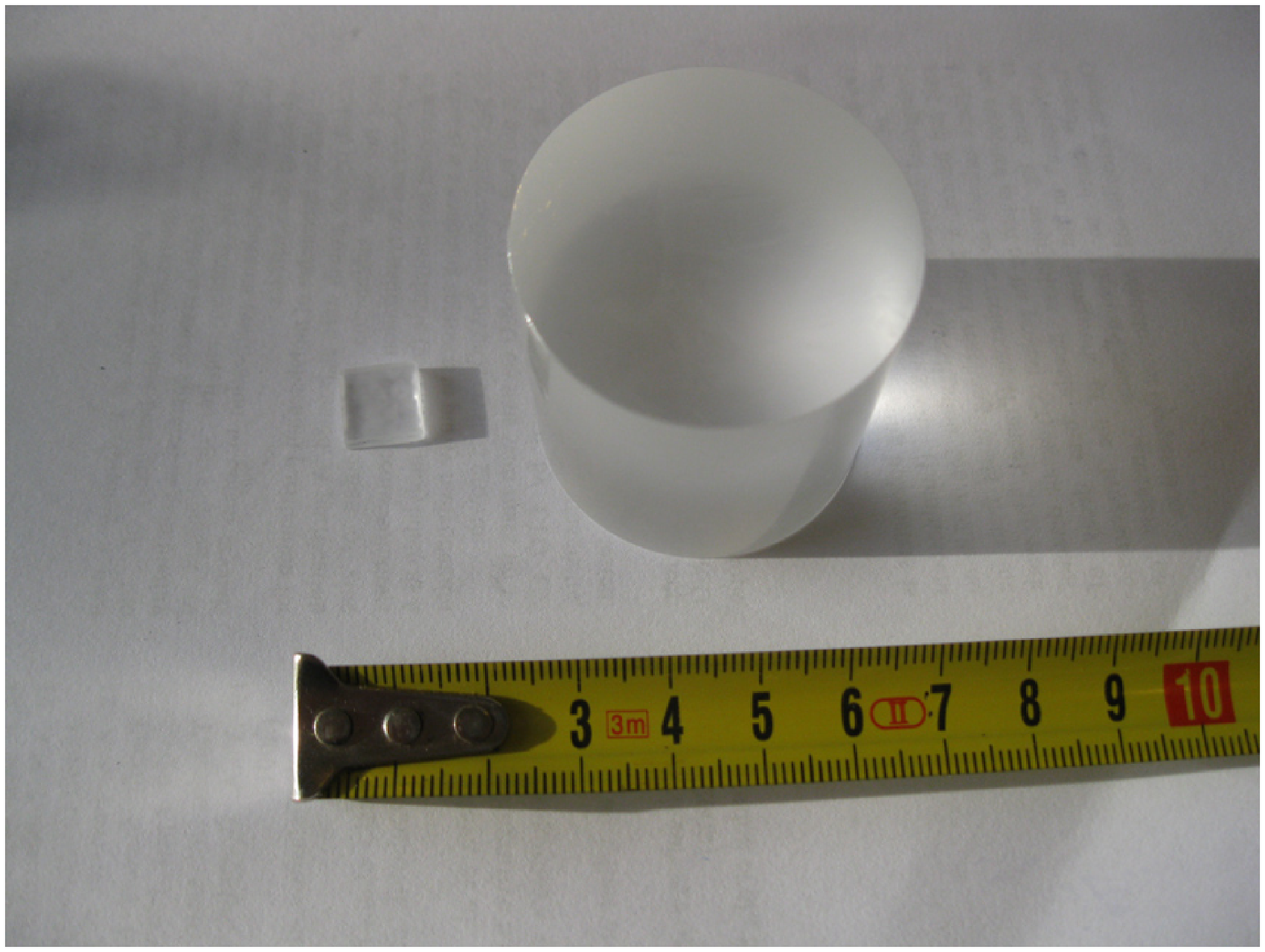}}
\caption{(Color online) Left: Li$_2$MoO$_4$ crystal boule.
Right: Li$_2$MoO$_4$ scintillation elements $10\times10\times 2$
mm (used for the luminescence measurements) and $\oslash 40\times
40$ mm (used for the cryogenic test) produced from the crystal
boule.}
 \label{crystals}
\end{center}
\end{figure*}

Lithium molybdate was studied as a promising material for cryogenic scintillating bolometer in 2010 by using a small 1.3 g crystal sample \cite{Barinova:2010}. The performance of a 33 g Li$_2$MoO$_4$ crystal working as a scintillating bolometer was presented in Ref.~\cite{Cardani:2013}. The first test of the material
radiopurity level was carried out with the help of ultra-low background HPGe gamma spectrometry at the Gran Sasso underground laboratory \cite{Barinova:2009}. Recently, Li$_2$MoO$_4$ crystals with diameter $\approx5$ cm were grown by Czochralski method \cite{Barinova:2014}.

The main properties of Li$_2$MoO$_4$ crystal scintillators are presented in Table~\ref{properties}.

The aim of our study was to develop medium volume Li$_2$MoO$_4$ scintillating bolometers -- 5 times larger than previous attempts~\cite{Cardani:2013} -- from deeply purified molybdenum by using the low thermal gradient Czochralski method for crystal growth, and to test the luminescence of the material and the scintillating bolometer performance at low temperatures. The aboveground cryogenic test has allowed also to estimate preliminarily the radiopurity level of the crystal and to ascertain that Li$_2$MoO$_4$ scintillating bolometers are excellent instruments for neutron detection thanks to the exothermic  $^6$Li(n,t)$\alpha$ reaction ($Q=4.78$~MeV).

\section{Development of Li$_2$MoO$_4$ crystal scintillators}

Deeply purified molybdenum oxide (MoO$_3$) \cite{Ber14} and commercial high purity lithium carbonate (Li$_2$CO$_3$, 99.99\%)
were used to prepare Li$_2$MoO$_4$ melt for crystal growth. The synthesis was carried out in a platinum crucible with a size of $\oslash70\times 130$ mm by using slow heating of the mixture ($\approx 50$ degree per hour). The compound was kept $5-10$ hours at a temperature of 450$^\circ$C.

A few Li$_2$MoO$_4$ crystal boules were grown in air atmosphere from the synthesized powder by the low thermal gradient Czochralski technique
\cite{Pavlyuk:1992,Borovlev:2001,Galashov:2010} in the platinum crucible of $\oslash70\times130$ mm. The temperature gradient was kept at the level of $\sim 1$ K/cm, which is one-two orders of magnitude lower in comparison to the conventional Czochralski method. Thanks to the low temperature gradient (and therefore absence of intensive evaporation of molybdenum oxide from the melt), the losses of the initial material did not exceed the level
of 0.5\% of the initial charge. This result is crucially important in view of the utilization of costly enriched material.  Output of crystal boules was at the level of up to 70\% of the initial charge. However, it should be noted that the efficiency of the growing process could be improved up to $80-90\%$ (already achieved for BGO~\cite{Grigoriev:2014}, cadmium tungstate~\cite{Belli:2010,Barabash:2011} and zinc molybdate~\cite{enr-above} crystal scintillators). The best quality crystals were obtained in growth conditions with a smooth slightly-convex crystallization front.

Optically clear defect-free Li$_2$MoO$_4$ crystals $25-55$ mm in diameter and $70-100$ mm in length with mass of $0.1-0.37$ kg were
grown (see Fig. \ref{crystals}, left). Two elements were cut from one of the boules for luminescence measurements ($10\times10\times2$ mm) and bolometric test ($\oslash 40\times40$ mm, see Fig. \ref{crystals}, right).

\section{Luminescence under X-ray excitation}

The luminescence of the ZnMoO$_4$ crystal sample ($10\times10\times 2$ mm) was investigated as a function of temperature between 8 K and 290 K under X-ray excitation. The sample was irradiated by X-rays from a tube with a rhenium anode (20 kV, 20 mA). Light from the crystal was detected in the visible region by a FEU-106 photomultiplier (sensitive in the wide wavelength region of $350-820$ nm) and in the near infrared region by a FEU-83
photomultiplier (with enhanced sensitivity in the near infrared wavelength region of $600-1200$ nm, with cooled photocathode). The measurements were carried out using a high-transmission monochromator MDR-2 (with the diffraction grating 600~mm$^{-1}$).

\begin{figure}
\begin{center}
\includegraphics[width=0.4\textwidth]{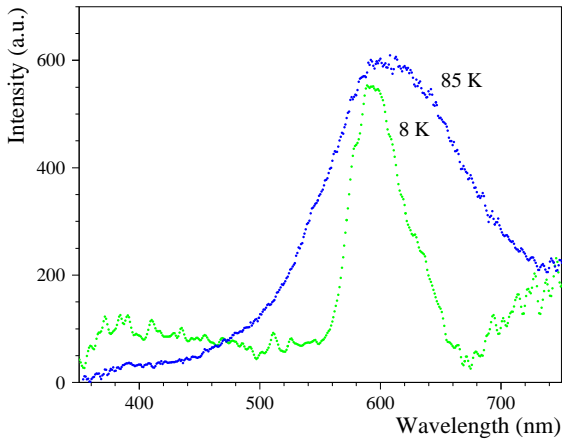}
\caption{(Color
online) Emission spectrum of Li$_2$MoO$_4$ crystal under X-ray
excitation at temperatures 8 K and 85 K.} \label{emission-sp}
\end{center}
\end{figure}

Emission spectra measured at temperatures of 8 K and 85 K, corrected for the spectral sensitivity of the registration system, are shown in Fig. \ref{emission-sp}. The most intensive emission band was observed in the spectra with maximum at  $\approx600$ nm at temperatures of 8 K and 85 K. The band at $\approx400$ nm is also observed at 8 K. The result is in a reasonable agreement with the data of previous measurements \cite{Barinova:2010}. Some shift of the emission maximum to longer wavelengths observed in the present study can be explained by a substantially higher purity of the initial materials (particularly of the molybdenum oxide) used in this study in comparison with the raw materials employed for the crystal growth in Ref.~\cite{Barinova:2010}.

\begin{figure}[b!]
\begin{center}
\includegraphics[width=0.4\textwidth]{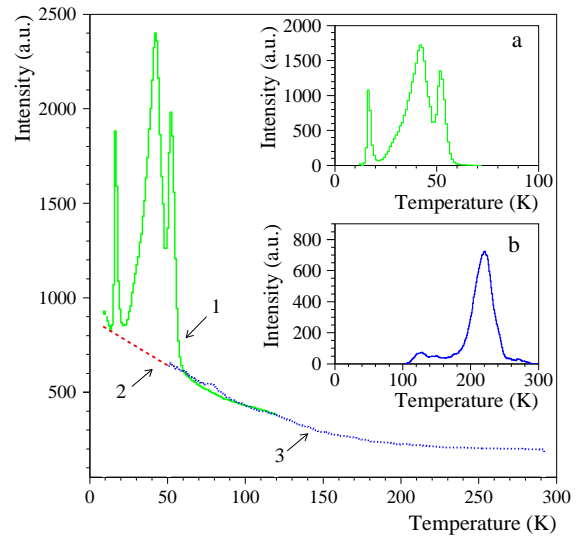}
\caption{(Color online) Temperature dependence of the luminescence
intensity under X-ray excitation. Experimental data (1) measured
under heating from liquid helium temperature contain a substantial
contribution of TSL.  Line (2) represents the behavior of the
luminescence intensity after subtraction of TSL. Curve (3) was
obtained under cooling from room temperature to $\approx50$ K.
(Inset a) Thermally stimulated luminescence of Li$_2$MoO$_4$ crystal
after X-ray excitation at 8 K. (Inset b) TSL of Li$_2$MoO$_4$ crystal
after X-ray excitation at 85 K.} \label{lum-int}
\end{center}
\end{figure}

The dependence of Li$_2$MoO$_4$ luminescence intensity on temperature was studied in the temperature interval $8-290$ K. There is a large contribution of thermo-stimulated luminescence (TSL) to the X-ray stimulated luminescence, in spite of a relatively low heating rate. To avoid the effect of TSL, we have subtracted the TSL contribution assuming a linear dependence of the luminescence on temperature. The obtained curve is presented in Fig. \ref{lum-int}. The luminescence intensity increases by a factor of 2 at 8 K in comparison with that at 85 K, and by a factor of 5 in comparison with that at room temperature. Such a behavior is rather different from the dependence observed in ZnMoO$_4$~\cite{Che13}, SrMoO$_4$~\cite{Jiang:2013} and PbMoO$_4$~\cite{Danevich:2010} crystal scintillators, where the intensity of luminescence drops down below the temperature range $60-120$ K. A similar temperature behavior (increase of scintillation efficiency with cooling down to liquid helium temperature) is exhibited by MgMoO$_4$~\cite{Mikhailik:2006},
CaMoO$_4$~\cite{Mikhailik:2010}, CdMoO$_4$~\cite{Mikhailik:2006}, and Li$_2$Zn$_2$(MoO$_4$)$_3$~\cite{Bashmakova:2009}.

\begin{figure}[t!]
\begin{center}
\includegraphics[width=0.4\textwidth]{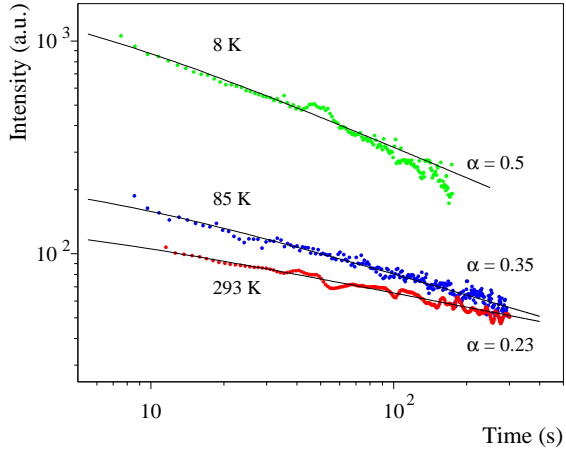}
\caption{(Color online) Time dependence of Li$_2$MoO$_4$ phosphorescence intensity after X-ray irradiation at temperatures 8~K, 85~K and 293~K. Fits
of the decay curves by hyperbolic function (see equation 1) are
shown by solid lines.}
\label{Phosphor}
\end{center}
\end{figure}

The TSL curves obtained with the Li$_2$MoO$_4$ sample after irradiation at 8 K and at 85 K are presented in the inset of Fig. \ref{lum-int}.

The long-term phosphorescence was measured at room temperature (293~K), at 8~K and 85~K (see Fig. \ref{Phosphor}). The phosphorescence decay curves can be approximated by the hyperbolic function:

\begin{equation}
I=\frac{I_{0}}{(1+a\times t)^{\alpha}}
\end{equation}

\noindent where $I$ is intensity of phosphorescence, $I_{0}$ is intensity of phosphorescence after the irradiation, $a$ is a coefficient which depends on the nature and properties of the traps in the samples, $\alpha$ is the degree of the hyperbolic function. A fit of the decay curves by a hyperbolic function gives $\alpha=0.5$, $\alpha=0.35$ and $\alpha=0.23$ for the phosphorescence after X-ray excitation at temperatures 8~K, 85~K and 293~K, respectively. The phosphorescence, as well as the intensive TSL, testify the presence of traps in the Li$_2$MoO$_4$ crystal. It should be noted that the hyperbolic behavior of the phosphorescence decay evidences the recombination character of luminescence in Li$_2$MoO$_4$.

\section{Bolometric tests in an aboveground laboratory}

\begin{figure}[b!]
\begin{center}
\includegraphics[width=0.37\textwidth]{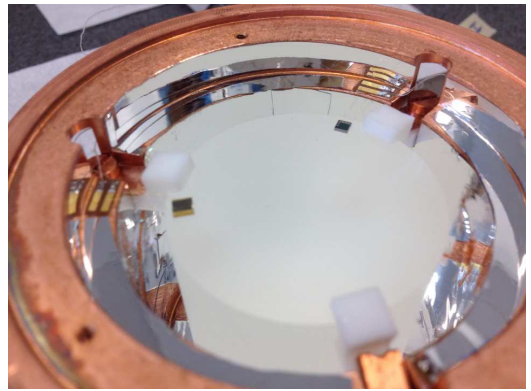}
\caption{(Color
online) Photograph of the upper face of the Li$_2$MoO$_4$ bolometer described in the text. The diameter of the crystal is 40~mm. The NTD Ge thermistor for thermal-signal read-out (attached at the left of the crystal face) and the heater for stabilization purpose are visible. Three S-shape PTFE elements keep the crystal in position in the copper holder.}
 \label{bolometer}
\end{center}
\end{figure}

The cylindrical crystal shown in Fig. \ref{crystals} (right panel), with a size of $\oslash 40\times 40$ mm and a mass of approximately $\sim 150$~g, was used to assemble a scintillating bolometer. The Li$_2$MoO$_4$ sample was fixed inside a cylindrical copper holder by using six PTFE elements. The lower face of the crystal sits on three PTFE L-shape elements which do not deform under crystal fixation. As shown in  Fig. \ref{bolometer}, three more PTFE S-shape elements rest on the upper face and exert a pressure on the crystal by a small deformation, fixing it at the copper support.  The Li$_2$MoO$_4$ was surrounded -- with the exception of the upper face -- by a reflecting foil (VM2000, VM2002 by 3M) to improve light collection.

The compound Li$_2$MoO$_4$  is slightly hygroscopic. This causes an affinity to atmospheric moisture which makes crystal surfaces whitish (see Fig.  \ref{bolometer}) due to exposure to air during detector assembly. No special precaution was taken for the mounting operation. The crystal (before assembly) and the detector (after assembly) were just kept under inert dry atmosphere. We tried to minimize the time required to install the bolometer in the cryostat.  The total exposure time to open atmosphere, including assembly and installation, was of the order of two days.

The Li$_2$MoO$_4$ crystal is equipped with a neutron trasmutation doped (NTD) Ge thermistor for the read-out of the heat signals. The mass of the Ge sensor is $\sim 50$~mg. It features a resistance of $\sim 500$ k$\Omega$ and a logarithmic sensitivity $A=-d\log(R)/d\log(T) \simeq 6.5$ at $\sim$20~mK. It is attached at the crystal surface by means of six epoxy glue spots and a 25 $\mu$m thick Mylar spacer, removed after the gluing operation. Four gold wires ($\oslash 25$~$\mu$m, length 15 mm) are ultrasonically bonded to the thermistor, providing both the electrical connection for the signals and the thermal link of the crystal to the copper heat sink. In addition, a heating element was glued at the crystal. It consists of a resistive meander of heavily-doped silicon with a low-mobility metallic behavior down to very low temperatures. The purpose of this heater is to deliver periodically a fixed amount of thermal energy in order to control and stabilize the thermal response of the bolometer according to the procedures described in Ref. \cite{heat1,heat2}. It was not used in the here-described aboveground tests, but it will be crucial for long-exposure underground runs foreseen for this scintillating bolometer.

\begin{figure}[t!]
\begin{center}
\includegraphics[width=0.4\textwidth]{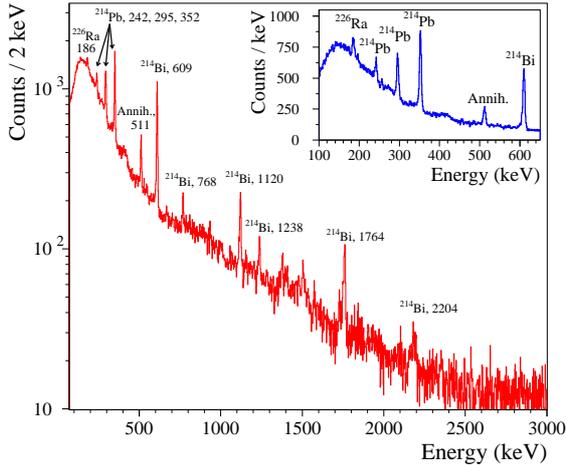}
\caption{(Color online) Energy spectrum measured in the Orsay aboveground set-up over 118 hours with the Li$_2$MoO$_4$ bolometer. Energies of $\gamma$ lines are in keV. In the inset, the low-energy part of the spectrum. The observed lines belong to the $^{238}$U radioactive chain and are all due to environmental radioactivity.}
\label{fig:spectrum}
\end{center}
\end{figure}

The choice of the light detector is very important, since the light yield of Li$_2$MoO$_4$ is expected to be remarkably lower~\cite{Cardani:2013} (by at least a factor 2) than that of other interesting materials to search for $0\nu2\beta$~\cite{Bha12,Gir10,Bee12,Art14,Bee12a,Bee12b,Bee12c,Che13,Ber14,Luc:2013,Cd1,Cd2}, as pointed out in Section~\ref{sec:intro}. We chose therefore a light detector with an advanced design~\cite{Cor2004} exhibiting a signal-to-noise ratio 3 -- 5 times higher than that achieved by the optical bolometers used so far in our tests with ZnMoO$_4$ detectors~\cite{Ten12}, although based on the same temperature sensor technology and read out by the same electronics. In the IAS laboratory (Orsay, France), where the detector was assembled and repeatedly tested, a baseline resolution of 50 eV (FWHM) was routinely achieved. The absorber of the advanced light detector is a germanium wafer with $\oslash 40$~mm and thickness 45 $\mu$m. We remind that germanium is an excellent radiation absorber from 1400 nm to X-ray. The Ge wafer suspension consists of 12 thin ($\oslash 6$~$\mu$m) low-heat-conduction superconductive wires. On the rear side a small NTD Ge sensor ($2\times0.4\times0.3$~mm) is glued, with a thermal link provided by a thin pure Ge slice with optimized design so that the heat crosses the sensor before the heat link. A low intensity $^{55}$Fe source facing the Ge wafer on the rear side is used for calibration. The optical bolometer is adapted to the larger heat bolometer cavity with a silver coated copper diaphragm. From previous tests with a similar optical configuration (performed at IAS with crystals of known luminescence like BGO, CaF$_2$(Eu) and CaWO$_4$) we can approximately estimate that the optical efficiency of the coupling between heat and light bolometers is $0.2 \pm 0.1$.

The scintillating bolometer was installed in the experimental vacuum of a high-power dilution refrigerator coupled to a pulse-tube cryostat \cite{Man14}, located aboveground at CSNSM (Orsay, France). In order to reduce pile-up effects, which can severely affect the aboveground measurements with slow large bolometers, the cryostat is surrounded by a massive low-activity lead shield with 10 cm maximum thickness. Following the approach described in Ref.~\cite{vib1,vib2}, the detector holder is suspended from the mixing chamber -- the coldest point of the dilution refrigerator -- by four stainless steel springs in order to damp acoustic vibrations which could generate excess noise through detector frictional heating. This arrangement works as a low-pass mechanical filter with a cut-off frequency at around 3 Hz. The thermal coupling to the mixing chamber is provided by two soft copper bands introducing a thermal time constant of about 2 s. The bolometric signals are read out and
 amplified by a room-temperature low-noise electronics, consisting of DC-coupled voltage-sensitive amplifiers with features similar to those described in Ref.~\cite{Arn02} and located inside a Faraday cage. Data were recorded in streaming mode by a 16 bit commercial ADC with 20 kHz sampling frequency.

The detector was operated at three different temperatures with the copper holder stabilized at 15.2 mK, 15.5 mK and 16.5 mK. The performance of the scintillating bolometer did not depend significantly on the base temperatures. The run at 15.5 mK had a duration of only 3.31 hours and has not be considered in data analysis. In the other two runs, we collected useful data for an overall duration of $\sim118$~hours. The resistances of the NTD thermistors at the operation point were in the range $1.2 - 1.6$~M$\Omega$ for both light detector and  Li$_2$MoO$_4$ bolometer, depending on the base temperature and bias level. Pulses induced by nuclear events and cosmic muons in the detectors were analyzed according to the optimum filter procedure, which consists in constructing and using a transfer function for the registered signal which optimizes the evaluation of the signal amplitude (which carries the information about the delivered energy), taking into account the pulse shape and the noise power spectrum~\cite{OF}.

Pulses of the Li$_2$MoO$_4$ detector had an amplitude of $40 - 60$~$\mu$V/MeV depending on the base temperature and operation point.  Rise and decay times were of $\sim10$~ms and $\sim75$~ms respectively, typical for large NTD-based bolometers. These parameters are computed as the time required by the pulse to vary from 10\% to 90\% (in case of rise time) and from 90\% to 30\% (in case of decay time) of its maximum amplitude. The intrinsic energy resolution was excellent, lying in the range $2-3$~keV  (computed as energy-converted FWHM baseline width after detector calibration with known-energy $\gamma$ events). These values are very promising as our experience shows that in underground runs -- where pulse pile-up does not affect $\gamma$ line energy resolution due to the low counting rate -- the energy resolution along the full spectrum is close to the intrinsic performance measured aboveground. In our aboveground spectra, the FWHM energy resolutions on $\gamma$ peaks was in the range $5 - 6$~keV at the 295.2~keV, 351.9~keV and 609.3~keV lines induced by  $^{214}$Po and $^{214}$Bi natural contamination. These values -- indicating anyway an excellent spectroscopic quality of the Li$_2$MoO$_4$ bolometer as appreciable in Fig.~\ref{fig:spectrum} -- are significantly worse than the intrinsic resolution because of the aforementioned pile-up effects (consider that the total counting rate above threshold was $\sim 2$~Hz) and detector response variations due to not fully recovered temperature instabilities.

The light detector developed at IAS confirmed its expected excellent performance. It was calibrated with an $^{55}$Fe X-ray source and exhibited an energy resolution of $\sim 300$~eV at the 5.9~keV line, probably limited by temperature instabilities. The FWHM baseline width, in the best run in terms of microphonic noise, was of the order of 73~eV and it has never been worse than about 80~eV, by far the best ever obtained with a light detector in the Orsay set-up and better than that achieved by the usual NTD-based light detectors developed for $0\nu2\beta$ decay search~\cite{Ten12,LD-GS}. Pulse rise and decay time, defined as for the heat channel, were as fast as  $\sim0.9$~ms and $\sim1.3$~ms respectively. This remarkable speed is obtained thanks to a careful design, described above, of the thermal couplings of the device. The pulse amplitude is also remarkable: in appropriate bias conditions it reached a value of 6.6 $\mu$V/keV, almost an order of magnitude higher than that observed in the devices described in Ref.~\cite{Ten12,LD-GS}.

\begin{figure}
\begin{center}
\includegraphics[width=0.4\textwidth]{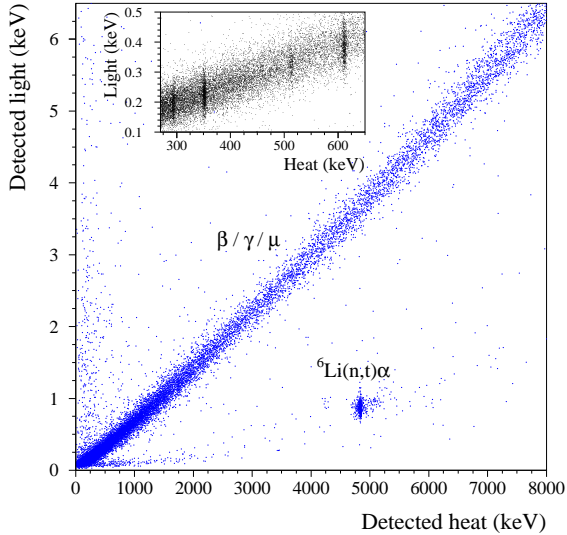}
\caption{(Color online) Scatter plot of the light signal amplitudes versus the heat signal amplitudes for a 25.3 h exposition of the Li$_2$MoO$_4$ scintillating bolometer to environmental radioactivity. The heat axis is calibrated using the known energy of the $\alpha$-triton peak. (Inset) Detail of the low energy part, where some of the $\gamma$ lines shown in Fig.~\ref{fig:spectrum} are appreciable as vertical clusters of points. The heat axis is calibrated using the identified $\gamma$ lines.}
\label{fig:scatter}
\end{center}
\end{figure}

Given the excellent performance of both the heat and the light channels, it was not difficult to collect high-quality heat-light scatter plots, as that shown in Fig.~\ref{fig:scatter}. This graph is built acquiring in coincidence the heat and the light pulses for the same event and reporting as a single point the two related amplitudes, determined with the optimum filter procedure and converted to energy using separate calibrations for the light and for the heat channels.

In the scatter plot, it is possible to recognize three main structures. A prominent fully-populated band contains $\beta$, $\gamma$ and cosmic muon events. We will refer to this structure as $\beta$ band. A detail of this band at low energies, shown in the inset, enlightens $\gamma$ characteristic lines observable also in the pulse amplitude spectrum (see Fig.~\ref{fig:spectrum}). The light yield related to this band is of the order of 0.7 keV/MeV at the energy of interest for $0\nu2\beta$, i.e. $\sim 3$~MeV. This value, which does not take into account the light collection efficiency, is $\sim 1.5$ higher than that reported in literature for a smaller previously operated Li$_2$MoO$_4$ sample~\cite{Cardani:2013}, which was of the order of $\sim 0.4$~keV/MeV, and not far from the value ($\sim 1 - 1.5$~keV/MeV ) obtained with the ZnMoO$_4$-based competing technology~\cite{Gir10,Bee12,Art14,Bee12a,Bee12b,Bee12c,Che13,Ber14,enr-above}.

\begin{figure}
\begin{center}
\includegraphics[width=0.4\textwidth]{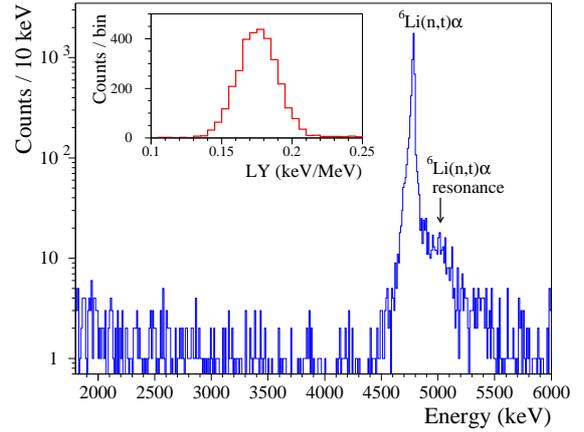}
\caption{(Color online) Energy spectrum of massive charged particles measured in the aboveground set-up over 118 hours with the Li$_2$MoO$_4$ scintillator, obtained by requiring that the light yield be less than 0.25 keV/MeV. The abscissa axis is calibrated using the known energy of the $\alpha$-triton peak. The inset shows the light-yield distribution in the neighborhood of the $\alpha$-triton peak. }
\label{fig:alpha}
\end{center}
\end{figure}

A second clear feature of the scatter plot is a cluster of points  
with a much lower light emission with respect to the main band. This class of events is observed in all Li-based (scintillating) bolometers~\cite{Cardani:2013,LIF-IAS-1,LIF-IAS-2,LIF-IAS-3}, and is due to neutron absorption. In fact, the isotope $^6$Li, which has a natural abundance of 7.5\%, has a very high cross section for thermal neutron capture (of the order of 940 barns). As mentioned above, this capture is followed by the emission of an $\alpha$ particle and a tritium nucleus, releasing a total kinetic energy of 4.78 MeV. The energy of the events in this cluster on a $\gamma$ calibrated scale (taking into account non-linearity) is about $\sim 5$\% higher than the nominal 4.78~MeV value, as commonly observed in scintillating bolometers when comparing $\beta$ and $\alpha$ thermal responses. The light yield at the $\alpha$-triton peak (of the order of 0.17 keV/MeV as appreciable in the inset of Fig.~\ref{fig:alpha}) is significantly lower than that of fast electrons. The corresponding quenching factor of the mixed $\alpha$-triton events with respect to $\beta$ events is about 23\% at $\sim 4.8$~MeV. As in the case of the $\beta$ band, the collected light energy is anyway  $\sim 1.5$~times higher than in Ref.~\cite{Cardani:2013}. About 0.24 MeV above the main $\alpha$-triton peak it is possible to observe a resonance enhancement corresponding to a well-known excited level of the compound nucleus $^7$Li$^*$. Although the neutron events do not correspond to a pure $\alpha$ signal and a small difference in quenching factor is expected between $\alpha$ particles and tritium nuclei~\cite{LIF-IAS-3}, Fig.~\ref{fig:scatter} shows clearly the excellent $\alpha$/$\beta$ separation power of our device.

A third structure observable in the scatter plot is a modestly populated band at low energies (extending up to $\sim 2$~MeV), with a much slower slope with respect to the main $\beta$ band. These events are due to nuclear recoils induced by fast neutrons. We remark that all the neutron-induced events detected by our device are due to the environmental background. No neutron source was used in these measurements. The potential of Li-based scintillating bolometers for both thermal and fast neutron detection is apparent and already discussed in literature~\cite{LIF-IAS-3}.

The sporadic points which belong to no structure in the scatter plot are related to high counting rate due to environmental radioactivity, and normally disappear in an underground measurement. Above the $\beta$ band these points are typically due to direct interaction of ionizing radiation in the light detector, which increases artificially the amplitude of the light signal. Below the $\beta$ band, we may have piled-up $\beta$-like events with a fake heat-to-light ratio, higher than that expected for such an interaction, as a consequence of the different time structure of light and heat signals. In addition, even if it is difficult to ascertain this point due to the low statistics,  there are some events compatible with an $\alpha$-like light-to-heat ratio between $\sim 2$ and $\sim 5$ MeV which could be due to energy-degraded $\alpha$'s associated to a surface contamination.

An aboveground measurement, because of the neutron background and the irreducible pile-up which prevents from obtaining clean scatter plots, is not suitable for an accurate determination of the crystal radiopurity, essential element to evaluate the sensitivity of $0\nu2\beta$ decay experiment based on the Mo compound under discussion. Nevertheless, the absence of internal $\alpha$ peaks in our exposure of about $\sim 18$~kg$\times$hour suggests an internal radioactivity due to uranium and thorium natural chains inferior to 1~mBq/kg. 

A typical contaminant of Li$_2$MoO$_4$ crystal is $^{40}$K~\cite{Barinova:2010}, because of the chemical affinity between lithium and potassium. However, the specific activity due to this radionuclide is very difficult to determine in an aboveground test, since it gives only counts inside the $\beta$ band and below $\sim 1.5$~MeV, where the counting rate induced by the environmental radioactivity is dominant. The $^{40}$K contamination cannot produce energy depositions extending up to the $0\nu2\beta$ decay spectral region. However, unresolved random coincidences between $^{40}$K-induced signals and events generated by other processes (especially the unavoidable $2\nu2\beta$ decay of $^{100}$Mo) can produce a pile-up spectrum populating the region of interest. It was shown however that this effect can be kept under control thanks to dedicated pile-up rejection algorithms~\cite{Che-pu1,Che-pu2} -- even in slow bolometers based on NTD Ge thermistors -- provided that the counting rate of the internal contamination does not exceed a few mBq/kg.  

Neutron-induced background is potentially harmful for a high sensitive $0\nu2\beta$ decay search. In particular, neutron capture on $^{7}$Li, the dominant isotope in natural lithium with an abundance of 92.5\%, produces the $\beta$-active isotope $^{8}$Li through an (n,$\gamma$) reaction with a cross section of 45.4 mbarn for thermal neutrons~\cite{Lyn1991,Hei1998}. The emitted $\gamma$ has an energy of 2.03 MeV and has therefore a good chance to escape a Li$_2$MoO$_4$ crystal with a volume of the order of $\sim 100$~cm$^3$ (typical for single bolometric modules in $0\nu2\beta$ decay search). The following $^{8}$Li $\beta$-decay ($\tau_{1/2} = 840$~ms~\cite{Ajz1984}) has a $Q$-value of 16.0~MeV and it can thus populate the region of interest of $0\nu2\beta$ decay of $^{100}$Mo. However, the $^{8}$Li decay occurs through a broad ($\Gamma$ = 1.5 MeV) $^{8}$Be excited state at $\sim 3$~MeV which disintegrates promptly into two $\alpha$ particles~\cite{Bha2006}. The absorbed $\beta$ electron and $\alpha$ particles, produced simultaneously within the detector time response, will give rise to single heat and light pulses. Therefore, the position of these $\alpha$/$\beta$ mixed events in the heat-light scatter plot will be below the $\beta$ band, where the signal is searched for, allowing for the rejection of this potential background.

\section{Conclusions}

Advanced quality lithium molybdate crystal scintillators were developed from deeply purified molybdenum oxide using the low thermal gradient Czochralski technique.

A weak luminescence of a Li$_2$MoO$_4$ crystal sample under X-ray irradiation was observed with a maximum emission at $\approx 600$ nm wavelength. The luminescence monotonically increases by a factor 5 with cooling from room to liquid helium temperature. Phosphorescence and intensive thermo-stimulated luminescence evidence the presence of traps due to defects in the crystal. Therefore there is still a margin to improve the quality of the material.

A few-day low-temperature test of a scintillating bolometer based on a Li$_2$MoO$_4$ cylindrical crystal -- with a size of $\oslash 40\times40$~mm --  was performed at $\sim 15$~mK in an aboveground pulse-tube cryostat housing a high-power dilution refrigerator in CSNSM (Orsay, France). The measurements demonstrated an excellent performance of the detector in terms of energy resolution and $\alpha$/$\beta$ separation power. The size of the scintillating bolometer is at least 5 times higher than that of any previously operated device based on Li$_2$MoO$_4$ and it is not far from the volume required for the single module of a large array to be used in a competitive $0\nu2\beta$ decay experiment.  There are also positive -- even  though very preliminary -- indications of a good radiopurity of the tested sample. A clear thermal neutron capture peak was observed from $^6$Li(n,t)$\alpha$ reaction.

The detector here described is under operation in a long underground run at the Gran Sasso National Laboratory in Italy, in order to study its radiopurity and the achievable performance in low counting rate conditions. Results will be presented in a dedicated work. An R\&D of Li$_2$MoO$_4$ crystal scintillators is in progress. The development of enriched Li$_2$$^{100}$MoO$_4$ scintillating bolometers is in progress too. We believe that the reported results are very encouraging in view of a future $0\nu2\beta$ decay experiment based on the technology here described, with the potential to deeply investigate the inverted hierarchy option of the neutrino mass ordering.

\section{Acknowledgments}

The development of Li$_2$MoO$_4$ scintillating bolometers was carried out within ISOTTA, a project receiving funds from the ASPERA 2$^{nd}$ Common Call dedicated to R\&D activities. The work was supported in part by the project ``Cryogenic detector to search for neutrinoless double beta decay of molybdenum'' in the framework of the Programme ``Dnipro'' based on Ukraine-France Agreement on Cultural, Scientific and Technological Cooperation. The group from the Institute for Nuclear Research (Kyiv, Ukraine) was supported in part by a Grant 9/14 of March 05, 2014 of the Space Research Program of the National Academy of Sciences of Ukraine. This work has been funded in part by the P2IO LabEx (ANR-10-LABX-0038) in the framework ``Investissements d'Avenir" (ANR-11-IDEX-0003-01) managed by the Agence Nationale de la Recherche (France).


\begin{thebibliography}{99}


\bibitem{Cre14} O.~Cremonesi, M.~Pavan, Challenges in Double Beta Decay, Adv. High En. Phys. 2014 (2014) 951432.
\bibitem{Sch13} B.~Schwingenheuer, Status and prospects of searches for neutrinoless double beta decay, Ann. Phys. 525 (2013) 269.
\bibitem{Ell12} S.R.~Elliott, Recent progress in double beta decay, Mod. Phys. Lett. A 27 (2012) 1230009.
\bibitem{Ver12} J.D.~Vergados, H.~Ejiri, F.~Simkovic, Theory of neutrinoless double-beta decay, Rep. Prog. Phys. 75 (2012) 106301.
\bibitem{Giu12} A.~Giuliani and A. Poves, Neutrinoless double-beta decay, Adv. High En. Phys. 2012 (2012) 857016.
\bibitem{Gom12} J.J.~Gomez-Cadenas et al., The search for neutrinoless double beta decay, Riv. Nuovo Cim. 35 (2012) 29.
\bibitem{Rod11} W.~Rodejohann, Neutrino-less double beta decay and particle physics, Int. J. Mod. Phys. E 20 (2011) 1833.
\bibitem{Qval} S. Rahaman et al., Q values of the $^{76}$Ge and $^{100}$Mo double-beta decays, Phys. Lett. B 662 (2008) 111.
\bibitem{AI} M.E. Wieser, J.R. De Laeter, Absolute isotopic composition of molybdenum and the solar abundances of the p-process nuclides $^{92,94}$Mo, Phys. Rev. C 75 (2007) 055802.
\bibitem{Fog2008} G.L. Fogli et al., Observables sensitive to absolute neutrino masses: A reappraisal after WMAP 3-year and first MINOS results, Phys. Rev. D 75 (2007) 053001.
\bibitem{Bar2012} J. Barea, J. Kotila, and F. Iachello, Limits on Neutrino Masses from Neutrinoless Double-$\beta$ Decay, Phys. Rev. Lett. 109 (2012) 042501.
\bibitem{Arn2014} R. Arnold et al., Search for neutrinoless double-beta decay of $^{100}$Mo with the NEMO-3 detector, Phys. Rev. D 89 (2014) 111101(R).
\bibitem{Pir2006} S. Pirro et al., Scintillating Double-Beta-Decay Bolometers, Phys. At. Nucl. 69 (2006) 2109.
\bibitem{Giu2012} A. Giuliani, Neutrino Physics with Low-Temperature Detectors, J. Low Temp. Phys. 167 (2012) 991.
\bibitem{Bha12} H. Bhang et al., AMoRE experiment: a search for neutrinoless double beta decay of $^{100}$Mo isotope with $^{40}$Ca$^{100}$MoO$_4$ cryogenic scintillating detector, J. Phys.: Conf. Ser. 375 (2012) 042023.
\bibitem{Kim14} G.B.~Kim et al., A CaMoO$_4$ crystal low temperature detector for the AMoRE neutrinoless double beta decay search, accepted by Adv. High En. Phys., Article ID 817530.
\bibitem{Gir10} L.~Gironi et al., Performance of ZnMoO$_4$ crystal as cryogenic scintillating bolometer to search for double beta decay of molybdenum, JINST 5 (2010) P11007.
\bibitem{Bee12} J.W.~Beeman et al., A next-generation neutrinoless double beta decay experiment based on ZnMoO$_4$ scintillating bolometers, Phys. Lett. B 710 (2012) 318.
\bibitem{Art14} D.R. Artusa et al., Exploring the neutrinoless double beta decay in the inverted neutrino hierarchy with bolometric detectors, Eur. Phys. J. C  74 (2014) 3096.
\bibitem{Bee12a} J.W.~Beeman et al., ZnMoO$_4$: A promising bolometer for neutrinoless double beta decay searches, Astropart. Phys. 35 (2012) 813.
\bibitem{Bee12b} J.W.~Beeman et al., An improved ZnMoO$_4$ scintillating bolometer for the search for neutrinoless double beta decay of $^{100}$Mo, J. Low Temp. Phys. 167 (2012) 1021.
\bibitem{Bee12c} J.W.~Beeman et al., Performances of a large mass ZnMoO$_4$ scintillating bolometer for a next generation 0$\nu$DBD experiment, Eur. Phys. J. C 72 (2012) 2142.
\bibitem{Che13} D.M.~Chernyak et al., Optical, luminescence and thermal properties of radiopure ZnMoO$_4$ crystals used in scintillating bolometers for double beta decay search, Nucl. Instr. Meth. Phys. Res. A 729 (2013) 856.
\bibitem{Ber14} L.~Berge et al., Purification of molybdenum, growth and characterization of medium volume ZnMoO$_4$ crystals for the LUMINEU program, JINST 9 (2014) P06004.
\bibitem{enr-above} A.S. Barabash et al., Enriched Zn$^{100}$MoO$_4$ scintillating bolometers to search for $0\nu2\beta$ decay of $^{100}$Mo with the LUMINEU experiment, Eur. Phys. J. C 74 (2014) 3133.
 \bibitem{Barinova:2010} O.P.~Barinova et al., First test of Li$_2$MoO$_4$ crystal as a cryogenic scintillating bolometer, Nucl. Instrum. Meth. Phys. Res. A 613 (2010) 54.
 \bibitem{Cardani:2013} L.~Cardani et al., Development of a Li$_2$MoO$_4$ scintillating bolometer for low background physics, JINST 8 (2013) P10002.
 \bibitem{Barinova:2009} O.P.~Barinova et al., Intrinsic radiopurity of a Li$_2$MoO$_4$ crystal, Nucl. Instrum. Meth. Phys. Res. A 607 (2009) 573.
 \bibitem{Barinova:2014} O.P.~Barinova et al., Properties of Li$_2$MoO$_4$ single crystals grown by Czochralski technique, J. Crystal Growth 401 (2014) 853.
 \bibitem{Denielou:1975} L.~Denielou et al., High-temperature calorimetric measurements: silver sulphate and alkali chromates, molybdates, and tungstates, J. Chem. Thermodynamics 7 (1975) 901.
 \bibitem{Tkhashokov:2009} N.I.~Tkhashokov et al., Li$_2$Mo$_4-$(NH$_{4})_{2}$MoO$_{4}-$H$_2$O System at
 25$^\circ$C, Russian Journal of Inorganic Chemistry 54 (2009) 1655 [Zhurnal Neorganicheskoi Khimii 54 (2009) 1732].
 \bibitem{Pavlyuk:1992} A.A.~Pavlyuk et al., Low Thermal Gradient technique and method for large oxide crystals growth from melt and flux, in Proc. of the APSAM-92, Asia Pacific Society for Advanced Materials, Shanghai, 26-29 April 1992, Institute of Materials Research, Tohoku University, Sendai, Japan, 1993, p. 164.
 \bibitem{Borovlev:2001} Yu.A.~Borovlev et al., Progress in growth of large sized BGO crystals by the low thermal gradient Czochralski technique, J. Cryst. Growth 229 (2001) 305.
 \bibitem{Galashov:2010} E.N.~Galashov et al., Growing of $^{106}$CdWO$_4$, ZnWO$_4$, and ZnMoO$_4$ scintillation crystals for rare events search by low thermal gradient Czochralski technique, Functional Materials 17 (2010) 504.
 \bibitem{Grigoriev:2014} D.N.~Grigoriev et al., Development of crystal scintillators for calorimetry in High Energy and Astroparticle
 Physics, JINST 9 (2014) C09004.
 \bibitem{Belli:2010} P.~Belli et al., Development of enriched $^{106}$CdWO$_4$ crystal scintillators to search for double $\beta$ decay processes in $^{106}$Cd, Nucl. Instr. Meth. Phys. Res. A 615 (2010) 301.
 \bibitem{Barabash:2011} A.S.~Barabash et al., Low background detector with enriched $^{116}$CdWO$_4$ crystal scintillators to search for double $\beta$ decay of $^{116}$Cd, JINST 6 (2011) P08011.
 \bibitem{Jiang:2013} H.~Jiang et al., Growth and Scintillation Characterizations of SrMoO$_4$ Single Crystals, J. Korean Phys. Soc. 63 (2013) 2018.
 \bibitem{Danevich:2010} F.A.~Danevich et al., Feasibility study of PbWO$_4$ and PbMoO$_4$ crystal scintillators for cryogenic rare events experiments,  Nucl. Instr. Meth. Phys. Res. A 622 (2010) 608.
 \bibitem{Mikhailik:2006} V.B.~Mikhailik and H.~Kraus, Cryogenic scintillators in searches for extremely rare events, J. Phys. D: Appl. Phys. 39 (2006) 1181.
 \bibitem{Mikhailik:2010} V.B.~Mikhailik and H.~Kraus, Performance of scintillation materials at cryogenic temperatures, Phys. Status Solidi B 247 (2010) 1583.
 \bibitem{Bashmakova:2009} N.V.~Bashmakova et al., Li$_2$Zn$_2$(MoO$_4$)$_3$ crystal as a potential detector for $^{100}$Mo $2\beta$-decay
 search, Functional Materials 16 (2009) 266.
\bibitem{heat1} A. Alessandrello et al., Methods for response stabilization in bolometers for rare decays, Nucl. Instr. Meth. Phys. Res. A 412 (1998) 454.
\bibitem{heat2} E. Andreotti et al., Production, characterization and selection of the heating elements for the response stabilization of the CUORE bolometers, Nucl. Instr. Meth. Phys. Res. A 664 (2012) 161.
\bibitem{Luc:2013} J.W. Beeman et al., Performances of a large mass ZnSe bolometer to search for rare events, JINST 8 (2013) P05021.
\bibitem{Cd1} L. Gironi et al., CdWO$_4$ bolometers for double beta decay search, Optical Materials 31 (2009) 1388.
\bibitem{Cd2} C. Arnaboldi et al., CdWO$_4$ scintillating bolometer for Double Beta Decay: Light and heat anticorrelation, light yield and quenching factors, Astropart. Phys. 34 (2010) 143.
\bibitem{Cor2004} N. Coron et al., Highly sensitive large-area bolometers for scintillation studies below 100 mK, Opt. Eng. 43 (2004) 1568.
\bibitem{Ten12} M.~Tenconi et al., Bolometric light detectors for Neutrinoless Double Beta Decay search, PoS (PhotoDet 2012) 072.
\bibitem{Man14} M. Mancuso et al., An aboveground pulse-tube-based bolometric test facility for the validation of the LUMINEU ZnMoO$_4$ crystals, J.  Low Temp. Phys. 176 (2014) 571.
\bibitem{vib1} S. Pirro et al., Vibrational and thermal noise reduction for cryogenic detectors, Nucl. Instr. Meth. Phys. Res. A 444 (2000) 331.
\bibitem{vib2} A. Alessandrello et al., A massive thermal detector for alpha and gamma spectroscopy, Nucl. Instr. Meth. Phys. Res. A 440 (2000) 397.
\bibitem{Arn02} C.~Arnaboldi et al.,  The Programmable Front-End System for CUORICINO, An Array of Large-Mass Bolometers, IEEE Trans. Nucl. Sci. 49 (2002) 2440.
\bibitem{OF} E. Gatti and P.F. Manfredi, Processing the signals from solid state detectors in elementary particle physics, Riv. Nuovo Cimento 9 (1986) 1.
\bibitem{LD-GS} J.W. Beeman et al., Characterization of bolometric light detectors for rare event searches, JINST 8 (2013) P07021.
\bibitem{LIF-IAS-1} P. de Marcillac et al., Characterization of a 2 g LiF bolometer, Nucl. Instrum. Meth. Phys. Res. A 337 (1993) 95.
\bibitem{LIF-IAS-2} C.S. Silver et al., Optimization of a $^6$LiF bolometric neutron detector,  Nucl. Instrum. Meth. Phys. Res. A 485 (2002) 615.
\bibitem{LIF-IAS-3} J. Gironnet et al., Neutron spectroscopy with $^6$LiF bolometers, AIP Conf. Proc. 1185 (2009) 751.
\bibitem{Che-pu1} D.M. Chernyak et al., Random coincidence of $2\nu2\beta$ decay events as a background source in bolometric $0\nu2\beta$ decay experiments, Eur. Phys. J. C 72 (2012) 1989.
\bibitem{Che-pu2} D.M. Chernyak et al., Rejection of randomly coinciding events in ZnMoO$_4$ scintillating bolometers, Eur. Phys. J. C 74 (2014) 2913. 
\bibitem{Lyn1991} J.E Lynn, E.T Jurney and S. Raman, Direct and valence neutron capture by $^{7}$Li, Phys. Rev. C 44 (1991) 764.
\bibitem{Hei1998} M. Heil et al., The (n,$\gamma$) cross section of $^{7}$Li,  ApJ 507 (1998) 997.
\bibitem{Ajz1984} F. Ajzenberg-Selove, Energy levels of light nuclei A = 5 -- 10, Nucl. Phys. A 413 (1984) 1.
\bibitem{Bha2006} M. Bhattacharya, E.G. Adelberger and H. E. Swanson, Precise study of the final-state continua in $^8$Li and $^8$B decays, Phys. Rev. C 73 (2006) 055802.


\end{thebibliography}
\end{document}